# Magnitude and spatial distribution control of the supercurrent in $Bi_2O_2Se$-based Josephson junction

*Jianghua Ying,* [†,‡] *Jiangbo He,* [†,‡] *Guang Yang,* [†,‡] *Mingli Liu,* [†,‡] *Zhaozheng Lyu,* [†,‡] *Xiang Zhang,* [†,‡] *Huaiyuan Liu,* [†,§] *Kui Zhao,* [†,‡] *Ruiyang Jiang,* [†,‡] *Zhongqing Ji,* [†,¶] *Jie Fan,* [†,¶] *Changli Yang,* [†] *Xiunian Jing,* [†,¶] *Guangtong Liu,* [†,¶] *Xuewei Cao,* [§] *Xuefeng Wang,* [#] *Li Lu,* [*,†,‡,¶,⊥,∥] *and Fanming Qu* [*,†,‡,¶,⊥]

[†] Beijing National Laboratory for Condensed Matter Physics, Institute of Physics, Chinese Academy of Sciences, Beijing 100190, China

[‡] School of Physical Sciences, University of Chinese Academy of Sciences, Beijing 100049, China

[§] School of Physics, Nankai University, Tianjin 300071, China

[¶] Songshan Lake Materials Laboratory, Dongguan, Guangdong 523808, China

[#] National Laboratory of Solid State Microstructures, Collaborative Innovation Center of Advanced Microstructures, School of Electronic Science and Engineering, Nanjing University, Nanjing 210093, China.

[⊥] CAS Center for Excellence in Topological Quantum Computation, University of Chinese Academy of Sciences, Beijing 100190, China

[∥] Beijing Academy of Quantum Information Sciences, Beijing 100193, China

ABSTRACT. Many proposals in exploring topological quantum computation are based on superconducting quantum devices constructed on materials with strong spin-orbit coupling (SOC). For these devices, a full control on both the magnitude and the spatial distribution of the supercurrent would be highly demanded, but has been elusive up to now. We constructed proximity-type Josephson junction on nanoplates of $Bi_2O_2Se$, a new emerging semiconductor with strong SOC. Through electrical gating, we show that the supercurrent can be fully turned ON and OFF, and its real-space pathways can be configured either through the bulk or along the edges. Our work demonstrates $Bi_2O_2Se$ as a promising platform for constructing multifunctional hybrid superconducting devices as well as for searching for topological superconductivity.

TEXT. Superconducting proximity effect (SPE) allows a normal metal to superconduct when placed adjacent to a superconductor. It plays a central role in the extensive applications of superconducting quantum devices. Recently, the search for topological superconductivity utilizes SPE as a primary mechanism to construct hybrid heterostructures, since intrinsic topological superconductors (TSCs) are scarce in nature[1-8]. Apart from being a novel topological phase of matter, TSCs afford platforms to hold Majorana zero modes (MZMs) which could be used for building fault-tolerant quantum computers[9-11]. On this fascinating blueprint, SPE has been realized in various topological materials and semiconductors with strong spin-orbit coupling (SOC), and TSCs and MZMs have been unveiled among these hybrid structures[2-4,12-16]. In spite of these rapid progresses, for the development of electrical tunable superconducting quantum information devices, control of both the magnitude and the spatial distribution of the supercurrent in nanoscale superconducting devices is a crucial ingredient. In superconductor-normal metal-superconductor (SNS) Josephson junctions, the control of the ON/OFF of the supercurrent through electrical gating (namely supercurrent transistor) has been demonstrated in quantum dots, nanowires/nanotubes, etc[17-19]. But the flow path of Cooper pairs is predefined by the geometry of the devices. Regarding regulating the spatial distribution of the supercurrent, in bulk-insulating SNS junctions, Cooper pairs flow naturally on the nontrivial two-dimensional (2D) surfaces of three-dimensional (3D) topological insulators and semimetals, or on the helical edges of 2D topological insulators[20-24]. However, the supercurrent is difficult to be turned OFF by gating due to the gapless nature in these topological materials. Efforts were also put on the monitor of the supercurrent distribution in ferromagnets and semiconductors such as InAs and graphene[25-31]. Nevertheless, simultaneous regulation of both the magnitude and the real-space pathways of the supercurrent in SNS junctions has been elusive, especially in semiconductors with strong SOC. From a fundamental viewpoint, SOC is essential to lift the spin degeneracy and is a prerequisite for most of the materials (trivial or nontrivial) explored so far to establish TSCs[1-4,6,13-15,20]. In this work, we address this issue and demonstrate the full control of the supercurrent through electrical gating in Josephson junction device on a new emerging star material with strong SOC - $Bi_2O_2Se$ nanoplates.

2D materials such as graphene, phosphorene, transition-metal dichalcogenides and topological insulators have attracted tremendous interest due to their promising applications in electronics, optoelectronics, etc[20,21,32-34]. Recently, the research on a new layered semiconductor $Bi_2O_2Se$ is booming thanks to its superior electronic properties such as the ambient-stability, an ultrahigh electron mobility (~2.8 × $10^5$ $cm^2$/V·s at 2 K), and a tunable bandgap[35-41]. The existence of strong SOC, suppression of backscattering, as well as high-performance field-effect transistors and optoelectronics, have been experimentally demonstrated in $Bi_2O_2Se$ thin films[42-46]. Coherent surface states have also been examined in $Bi_2O_2Se$ nanowires[47], and a large Rashba SOC was further predicted to be present on the polar or nonpolar surfaces due to the intrinsic band bending[40]. These surface states reside partially in the bulk band gap, leaving appropriate space for multifunctional electrical tuning. Given these intriguing properties, $Bi_2O_2Se$ provides opportunities to manipulate SPE and a promising platform to engineer TSCs.

In this work, we study Josephson junction constructed on a $Bi_2O_2Se$ nanoplate and demonstrate the full control of the magnitude and the spatial distribution of the supercurrent through electrical gating. By employing superconducting interferometry techniques, we found that the supercurrent can be fully tuned ON and OFF, and meanwhile, the supercurrent flow can be configured either through the bulk or only along the edges. Two alternative mechanisms of the bulk to edge supercurrent transition are presented.

$Bi_2O_2Se$ nanoplates were synthesized by means of chemical vapor deposition in a horizontal tube furnace[35,48]. Figure 1a shows an optical microscope image of the as-grown nanoplates, viewed perpendicular to the mica substrate. The nanoplates take a rectangular shape, and either lie down on the substrate or grow vertically as indicated by the red arrows in Fig. 1a. The standing $Bi_2O_2Se$ nanoplates can be easily transferred onto $SiO_2$/Si substrates using a purely mechanical way instead of a wet transferring process mediated by polymethylmethacrylate, to avoid possible contaminations[42,46]. Electrode patterns were fabricated by electron-beam lithography, and Ti/Al (5 nm/65 nm) contacts were deposited by electron-beam evaporation. Note that a soft plasma cleaning of the surface of $Bi_2O_2Se$ was performed prior to the metal deposition to improve the interface transparency. Figure 1b presents the false-colored scanning electron microscope (SEM) image of the Josephson junction device studied in this work. The Ti/Al electrodes (yellow) have a width of 560 nm each and a separation of 300 nm. The width of the $Bi_2O_2Se$ nanoplate (dark blue) is $W$=4.6 μm and a thickness of 15 nm is confirmed by atomic force microscopy (see Supplementary information). Quasi-four terminal measurement, as depicted in Fig. 1b, was performed with standard low-frequency lock-in techniques in a dilution refrigerator at a base temperature of ~10 mK. A gate voltage $V_g$ was applied through the degenerately doped Si substrate covered with 300 nm thick $SiO_2$.

For SNS Josephson junctions, superconducting interferometry is a convenient and powerful technique to discriminate the spatial distribution of the supercurrent[20,21,27]. The supercurrent-density profile can be extracted from the perpendicular magnetic field dependence of the critical current, $I_c$ vs $B_z$, through the Dynes-Fulton approach[49]. Figure 1c illustrates the correspondence between the interference patterns (right column) and the supercurrent density as a function of position, $J_s(x)$ (left column). For a spatially uniform supercurrent (top row), which is usually the case for a bulk conducting junction, $I_c(B_z)$ presents a one-slit Fraunhofer-like pattern following the form $|\sin(\pi\Phi/\Phi_0)/(\pi\Phi/\Phi_0)|$, where $\Phi=L_{eff}WB_z$ is the flux, $L_{eff}$ and $W$ are the effective length and width of the junction, respectively, and $\Phi_0=h/2e$ is the flux quantum ($h$ is the Planck constant, $e$ the elementary charge). The central lobe of $I_c(B_z)$ has a width of $2\Phi_0$ and the side lobes of $\Phi_0$. In addition, the height of the lobes shows an overall $1/|B_z|$ decay. When the supercurrent flows along the two edges (bottom row), the single junction mimics a superconducting quantum interference device (SQUID) and presents a two-slit interference pattern with the form $|\cos(\pi\Phi/\Phi_0)|$. The lobes in this case have a uniform width of $\Phi_0$ and a weak overall decay (the decrease mainly results from the suppression of the superconductivity in the contacts and a finite width of the edge modes)[50]. In the intermediate regime (middle row), bulk and edge supercurrent coexist, and $I_c(B_z)$ exhibits a one-slit to two-slit transition. Generally speaking, $I_c(B_z)$ and $J_s(x)$ can be extracted quantitatively from each other through integration or Fourier transform[49].

We first present the gate tuning of the magnitude of the supercurrent between the ON and OFF states. Figures 2a and 2b show the differential resistance $dV/dI \equiv V_{ac}/I_{ac}$ as a function of gate voltage $V_g$ and dc bias current $I_{dc}$. To reach a high contrast and a clear view of the color map, we divided the full $V_g$ range of -10 V to 60 V into these two subranges and plotted them in two different color scales. The $dV/dI$ peaks, as indicated by the white arrows, correspond to the critical current $I_c$, separating the zero-resistance state (uniform blue color) at low $|I_{dc}|$ from the normal state at high $|I_{dc}|$. Two typical line cuts are plotted in Fig. 2d, taken at $V_g$=20 V (blue) and 40 V (red), respectively. Figure 2c shows the gate voltage dependence of the normal-state resistance $R_N$ and $I_c$ extracted from Figs. 2a and 2b. When sweeping $V_g$ from 60 V to -10 V, $R_N$ increases monotonically towards infinite (~50 kΩ at $V_g$=-20 V, dada not shown), which is expected as the electrons are depleted. Meanwhile, $I_c$ decreases also monotonically from 223 nA at $V_g$=60 V to 0 nA around $V_g$=-7 V. Therefore, the supercurrent can be fully turned ON/OFF by gating. Note that the black arrows in Figs. 2a and 2c indicate the upturn of $I_c$ around $V_g$=15 V which will be discussed later. The $I_cR_N$ product is about 8 µV at $V_g$ = 60 V, with an energy scale much smaller than the superconducting gap of Al, indicating a low-transparency interface.

Next, we switch to the superconducting interference and disentangle the bulk and edge supercurrent. To do so, a perpendicular magnetic field $B_z$ was applied to the junction. Figure 3a shows the differential resistance $dV/dI$ as a function of both $B_z$ and $I_{dc}$ at $V_g$=60, 40, 20, and 0 V, respectively. The whitish envelope characterizes the critical current $I_c$ and separates the superconducting and normal states. At $V_g$=60 V, $I_c$ presents a one-slit Fraunhofer-like pattern, with a central lobe of width roughly twice of the side lobes and also a global fast decay. The average period of $\Delta B_z \approx$5.1 Gauss agrees with the dimension of the junction – $\Phi_0/L_{eff}W$=5.2 Gauss, where $L_{eff}$=860 nm is the effective length of the junction considering the flux focusing effect of the contacts. The two dashed lines illustrate the positions of the first nodes. At $V_g$=40 V, not much changes but a smaller $I_c$ as the carrier density reduces. When $V_g$ decreases further to 20 V, the first lobe shrinks in both width and amplitude, and the nodes move towards zero $B_z$. Moreover, the overall decay of $I_c$ slows down, indicating a transition from bulk to edge supercurrent. Finally, at $V_g$=0 V, a clear two-slit SQUID pattern appears while the nodes move inwards further, forming equally spaced, weakly decayed lobes, which signals edge-mode superconductivity.

The analysis above can also be investigated by using the Dynes-Futon approach to extract the supercurrent-density profile. Figure 3b displays the position ($x$) dependence of $J_s$ at different $V_g$, obtained from the Fourier transform of the $I_c$ vs $B_z$ curves retrieved from Fig. 3a accordingly. Note that $x$=0 corresponds to the center of the junction. The bulk-dominated supercurrent at $V_g$=60 and 40 V, the bulk to edge transition at $V_g$=20 V, and the edge-dominated supercurrent at $V_g$=0 V can be immediately recognized. Therefore, through electrical gating, we realized the configuration of the supercurrent pathways.

We now discuss the underlying mechanism of the gate tuning of the supercurrent distribution. Although the Fraunhofer to SQUID evolution and the supercurrent-density profiles demonstrate the "bulk" to edge supercurrent transition, it is ambiguous that if the "bulk" is the 3D bulk or alternatively, the top and bottom surfaces. These two scenarios arise from the Se or Bi termination-sensitive surface states in $Bi_2O_2Se$.

For Se-terminated (001) surface, Fig. 4a sketches the band structure (not to scale) around Γ point, including the spin-split subbands (blue) close to the valence band. Regarding the fact that when sweeping $V_g$ from 60 V to -10 V, the normal state resistance increases monotonically towards infinite, the conduction band of the (3D) bulk undertakes the electron transport, as indicated by the blue shaded region. However, the bulk to edge supercurrent transition suggests the existence of edge states, presumably due to downward band bending at the edges[26,40,51]. Figure 4b displays such band bending and the formation of the edge states (at energy $E'$), as illustrated by the red dashed line in Fig. 4a. Consequently, edge supercurrent dominates when the Fermi level $E_F$ sits between $E'$ and the bottom of the conduction band $E_C$ (bottom panel of Fig. 4c). The upturn of the critical current around $V_g$=15 V, as guided by the black arrows in Figs. 2a and 2c, corresponds to the onset of the 3D bulk supercurrent when $E_F>E_C$ (top panel of Fig. 4c), consistent with the observation of finite $J_s$ in the bulk at $V_g$=20 V (Fig. 3).

The scenario for Bi-terminated top/bottom surface is completely different[40]. Figure 4d sketches the band structure (not to scale) around Γ point, including the two spin-split subbands (blue) on (001) surface. According to Ref. 40, the separation between the subband bottoms is $E_2-E_1 \approx 0.34$ eV. We assign $V_g$=-10 V to correspond to a Fermi level position just below $E_1$, since the junction tends to be insulating. Thus, the Fermi level position $E_F(V_g) = \pi\hbar^2 n / 2m^* + E_1 = \pi\hbar^2 \varepsilon (V_g + 10 \text{ V}) / 2edm^* + E_1$, where $n$ is the 2D electron density induced by the gate, a factor of 2 denotes an equal footing of the top and bottom surfaces, $\varepsilon$ the dielectric constant of SiO$_2$, $d$=300 nm the thickness of SiO$_2$, $m^*$=0.14$m_0$ the effective electron mass of Bi$_2$O$_2$Se ($m_0$ is the electron mass)[35]. At $V_g$=60 V, $E_F$ sits ~0.04 eV above $E_1$, far below $E_2$ and the conduction band $E_C$. Hence, in the gate-voltage range studied, as indicated by the blue shaded region in Fig. 4d, electrons are only induced on the surfaces of the Bi$_2$O$_2$Se nanoplate, but not in the 3D bulk. Consequently, the "bulk" supercurrent mentioned above flows actually through the top and bottom surfaces. We would like to note that there can be a difference between these two surfaces due to the gate being at the bottom and the contacts being on the top.

Similarly, the bulk to edge transition indicates a slightly different $E_1$ (onset energy) between the top/bottom surfaces and the side surfaces. Under this assumption, Fig. 4e illustrates such effective band bending and the formation of the surface subbands. When $E_F$ sits above $E_1$, the top and bottom surfaces dominate the supercurrent with relatively small contributions from the side surfaces considering the small thickness of 15 nm (top panel of Fig. 4f). But within the narrow energy window between $E'_1$ and $E_1$, the side surfaces dominate (bottom panel of Fig. 4f). Again, the upturn of $I_c$ around $V_g$=15 V implies the onset of the supercurrent on the top and bottom surfaces.

For both Se and Bi terminations, however, the full width at half maximum (FWHM) of the edge-supercurrent density peak, ~570 nm (see bottom right panel in Fig. 3b), is much larger than the typical thickness of surface states, say several nm[40,52]. This can be explained by the extension of the 2D electron waves from the edge to the 3D bulk for Se-terminated case (from side surfaces to the top and bottom surfaces for Bi-terminated case) with an order of the Fermi wave length, as observed in graphene[27]. Such explanation is further corroborated by the decay of the supercurrent density from the edge towards the center of the junction (see Supplementary information). Notably, these two scenarios are very interesting, though, the current study could not differentiate the Se- or Bi-terminated surfaces, and alternative interpretations for the edge supercurrent may also apply. Therefore, detailed studies are required to nail down more exquisite physical pictures. Note that the crystal structure, the thickness, and the edge characteristics of $Bi_2O_2Se$ nanoplates are different from gapped graphene, so that the mechanism for the transition of the supercurrent distribution belongs to different scenarios[28].

In summary, we fabricated Josephson junction device on a $Bi_2O_2Se$ nanoplate and realized superconducting proximity effect. Through electrical gating, the supercurrent can be fully turned ON and OFF, and simultaneously, the supercurrent spatial distribution can be configured in the bulk or along the edges. The termination-sensitive surface states render plentiful mechanisms and various possible configurations, and leave extensive room for multifunctional electrical tuning. Considering the remarkable properties and the strong spin-orbit coupling, $Bi_2O_2Se$ is a promising platform for studying numerous novel phenomena, such as constructing gate controllable hybrid superconducting devices[18], supercurrent field-effect transistors[19], nano-scale SQUID[53], superconducting optoelectronics[54], Josephson laser[55], Cooper-pair beam splitter[56], and engineering topological superconductors[2,3,6].

## ASSOCIATED CONTENT

**Supporting Information.** The Supporting Information which contains additional figures and analysis is available free of charge on the ACS Publications website.

## AUTHOR INFORMATION

**Corresponding Authors**

* Emails: lilu@iphy.ac.cn, fanmingqu@iphy.ac.cn.

**Author Contributions**

L.L. and F.Q. supervised the overall research. J.Y. and R.J. synthesized the material. J.Y. fabricated the device and carried out the measurement with the assistance from J.H., G.Y., M.L., Z.L., X.Z., H.L., K.Z., R.J., Z.J., J.F., C.Y., X.J., G.L., L.L. and F.Q.

All authors participated in the analysis of the data. F.Q., J.Y., and L.L. wrote the paper under constructive discussions with all other authors.

**Notes**

The authors declare no competing financial interests.

ACKNOWLEDGMENT

We would like to thank Hailin Peng and Ning Kang for fruitful discussions. This work was supported by the National Basic Research Program of China from the MOST grants 2017YFA0304700, 2016YFA0300601, and 2015CB921402, by the NSF China grants 11527806, 91221203, 11174357, 91421303, 11774405 and 61822403, by the Strategic Priority Research Program B of Chinese Academy of Sciences, Grants No. XDB28000000 and XDB07010100, by the Beijing Municipal Science & Technology Commission, China, Grant No. Z191100007219008, by the Open Research Fund from State Key Laboratory of High Performance Computing of China, and by Beijing Academy of Quantum Information Sciences, Grant No. Y18G08.

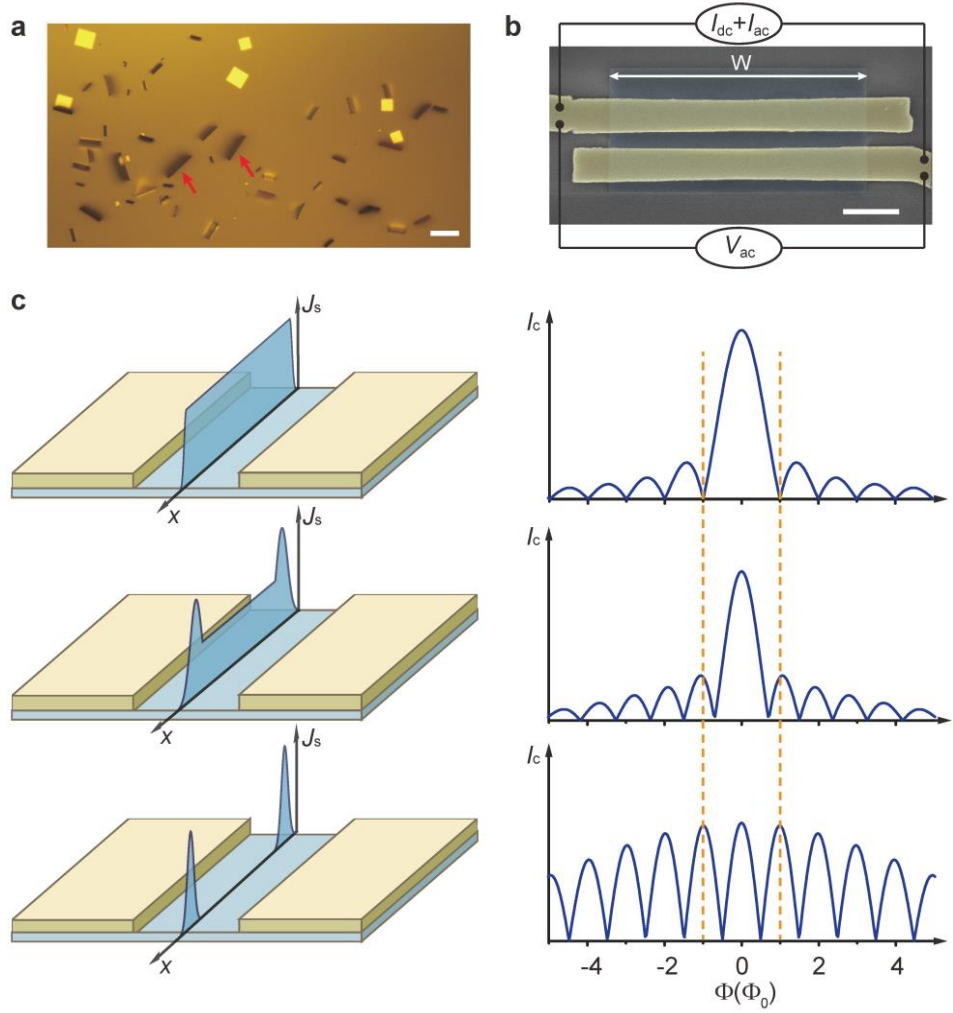

**Figure 1. Device and superconducting interferometry. a**, Optical microscope image of the as-grown $Bi_2O_2Se$ nanoplates, with a rectangular shape, either lying or standing (as indicated by the red arrows) on the mica substrate. Scale bar: 10 μm. **b**, False-colored scanning electron microscope image of the Josephson junction studied. Scale bar: 1 μm. Two Ti/Al electrodes (yellow) contact the $Bi_2O_2Se$ nanoplate (dark blue) forming a proximity-type Josephson junction. The circuit displays the quasi-four terminal measurement configurations. **c**, Correspondence between the supercurrent-density profile, $J_s(x)$, in a Josephson junction (left column) and the superconducting interference pattern, i.e., the dependence of critical current $I_c$ on magnetic flux Φ in the junction area (right column). The two dashed lines indicate the shift of the positions of the first nodes relative to the one-slit Fraunhofer-like pattern (top right).

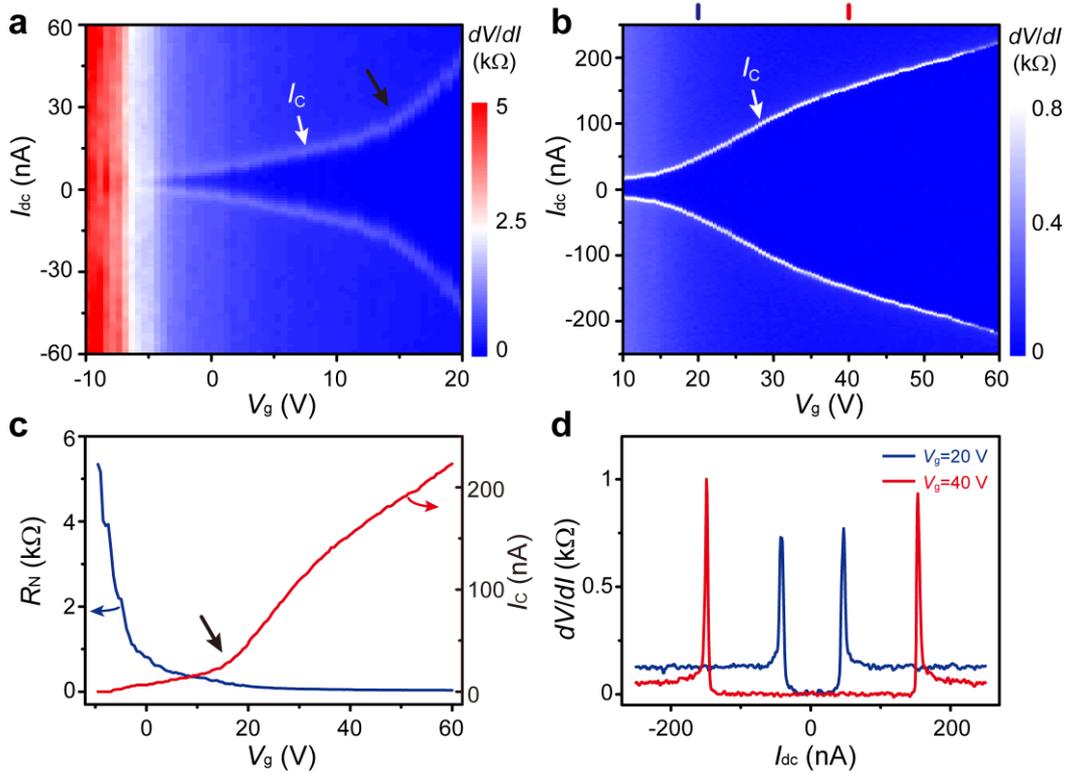

**Figure 2. Gate tunable ON and OFF states of the supercurrent. a** and **b**, Differential resistance $dV/dI$ as a function of both gate voltage $V_g$ and dc bias current $I_{dc}$, in two gate-voltage ranges for clarity. The white arrows mark the critical current $I_c$. **c**, $V_g$ dependent normal state resistance $R_N$ and $I_c$ extracted from **a** and **b**. **d**, Typical line cuts at $V_g$=40 V (red) and 20 V (blue) taken from **b** as marked by the red and blue bars, showing the superconducting and normal states at low and high bias current, respectively. The black arrows in **a** and **c** guide the upturn of $I_c$ around $V_g$=15 V.

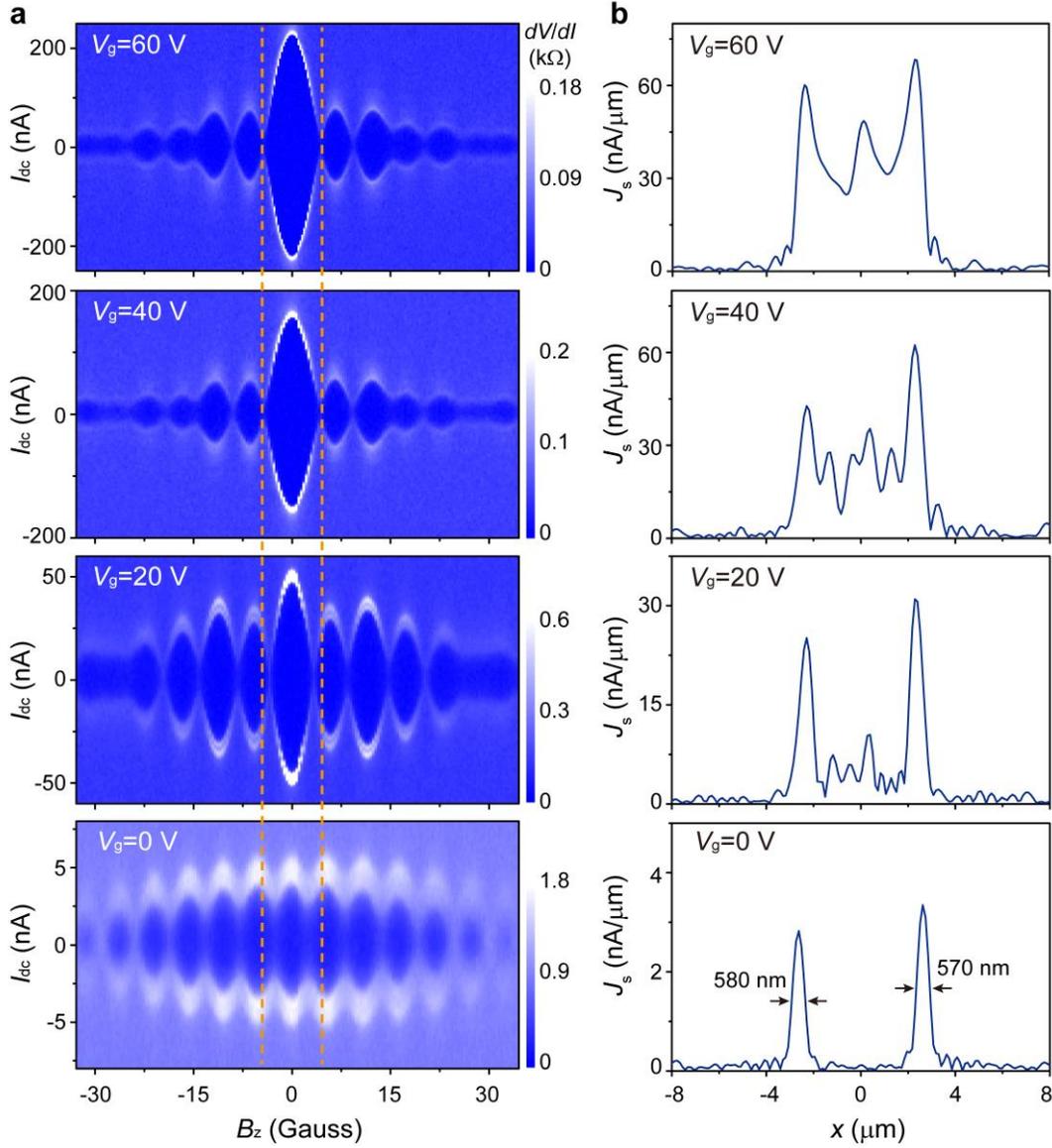

**Figure 3. Bulk to edge supercurrent transition. a**, Superconducting interference patterns - differential resistance $dV/dI$ as a function of perpendicular magnetic field $B_z$ and dc bias current $I_{dc}$, at $V_g$=60, 40, 20 and 0 V, respectively. The two dashed lines illustrate the shift of the first nodes. **b**, Supercurrent-density $J_s(x)$ retrieved from **a** at different gate voltages accordingly, through the Dynes-Fulton approach. $x$=0 corresponds to the center of the junction. The FWHM of the left and right peak when $V_g$=0 V is 580 nm and 570 nm, respectively.

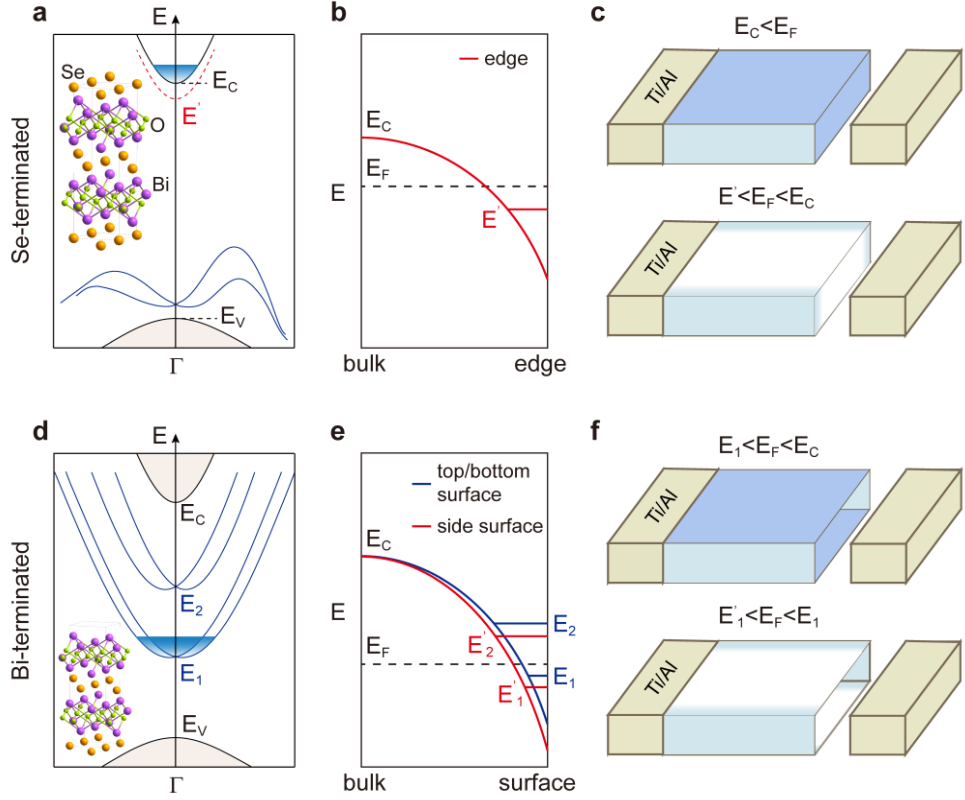

**Figure 4. Mechanisms of the bulk to edge supercurrent transition. a, d**, Sketches of the band structure of the Se-terminated (**a**) and Bi-terminated (**d**) (001) surface around Γ point[40]. The spin-split surface states are represented by the blue lines. The blue shaded regions indicate the range corresponding to the gate-voltage applied in this study. The insets present the Se- and Bi-terminated (001) surface, respectively. **b**, Effective band bending at the edges of the $Bi_2O_2Se$ nanoplate for Se-terminated top/bottom surfaces and the formation of the edge states. **c**, Illustration of the bulk-dominated (top panel) and the edge-dominated (bottom panel) supercurrent for Se-terminated case, with the blue color indicating the supercurrent pathways. For clarity, the right Ti/Al contacts are moved apart. **e**, Effective band bending at the surfaces for Bi-terminated case and the formation of the surface subbands which are assumed to be slightly lower for the side surface (red) than that for the top/bottom surface (blue). **f**, Illustration of the top and bottom surfaces dominated (top panel) and the side surfaces dominated (bottom panel) supercurrent for Bi-terminated case.

Supporting Information for

Magnitude and spatial distribution control of the supercurrent in $Bi_2O_2Se$-based Josephson junction

*Jianghua Ying,* [†,‡] *Jiangbo He,* [†,‡] *Guang Yang,* [†,‡] *Mingli Liu,* [†,‡] *Zhaozheng Lyu,* [†,‡] *Xiang Zhang,* [†,‡] *Huaiyuan Liu,* [†,§] *Kui Zhao,* [†,‡] *Ruiyang Jiang,* [†,‡] *Zhongqing Ji,* [†,¶] *Jie Fan,* [†,¶] *Changli Yang,* [†] *Xiunian Jing,* [†,¶] *Guangtong Liu,* [†,¶] *Xuewei Cao,* [§] *Xuefeng Wang,* [#] *Li Lu,* [\*,†,‡,¶,⊥,∥] *and Fanming Qu* [\*,†,‡,¶,⊥]

[†] Beijing National Laboratory for Condensed Matter Physics, Institute of Physics, Chinese Academy of Sciences, Beijing 100190, China

[‡] School of Physical Sciences, University of Chinese Academy of Sciences, Beijing 100049, China

[§] School of Physics, Nankai University, Tianjin 300071, China

[¶] Songshan Lake Materials Laboratory, Dongguan, Guangdong 523808, China

[#] National Laboratory of Solid State Microstructures, Collaborative Innovation Center of Advanced Microstructures, School of Electronic Science and Engineering, Nanjing University, Nanjing 210093, China.

[⊥] CAS Center for Excellence in Topological Quantum Computation, University of Chinese Academy of Sciences, Beijing 100190, China

[∥] Beijing Academy of Quantum Information Sciences, Beijing 100193, China

[\*] Corresponding authors: lilu@iphy.ac.cn, fanmingqu@iphy.ac.cn.

## Contents

1. **Synthesis of $Bi_2O_2Se$ nanoplates**
2. **Device fabrication**
3. **Low temperature transport measurement**
4. **Characterization of $Bi_2O_2Se$**
5. **Extraction of the supercurrent-density profile**
6. **Comparison between different devices**

1. **Synthesis of Bi$_2$O$_2$Se nanoplates**

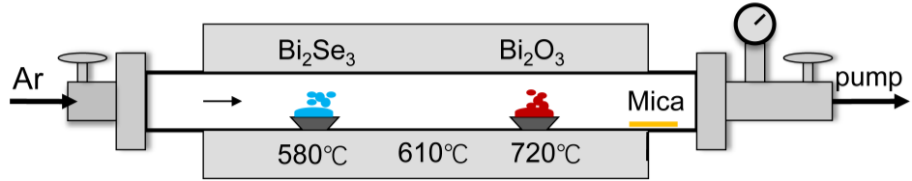

**Figure S1. Synthesis of Bi$_2$O$_2$Se nanoplates.** Bi$_2$O$_2$Se nanoplates were synthesized in a three-zone horizontal tube furnace[1,2]. Bi$_2$Se$_3$ powder (0.6 g) was put at the first hot zone (upstream), and Bi$_2$O$_3$ powder (0.8 g) was put at the third hot zone, as the sources. Freshly cleaved mica substrates were put at the downstream with a temperature gradient. During growth, a 200 sccm argon carrier gas was applied while keeping the pressure between 50 – 200 Torr. The temperature for the first and second zones were kept typically at 580 °C and 610 °C, respectively, and the third zone was regulated between 620 °C and 720 °C. The growth time ranges from 10 to 60 min. Note that a rapid cool-down is preferred after growth.

**2. Device fabrication.** Bi$_2$O$_2$Se nanoplates were mechanically transferred from mica onto a degenerately doped Si substrate with a 300 nm thick SiO$_2$ layer used for applying a back-gate voltage. This was performed just by touching the mica substrate with the Si substrate face to face, and the standing Bi$_2$O$_2$Se nanoplates were relocated onto the Si target. Afterwards, electrode patterns were fabricated by electron-beam lithography using pre-defined markers, and Ti/Al (5 nm/65 nm) contacts were deposited by electron-beam evaporation. Prior to the metal deposition, an in-situ soft Ar plasma cleaning at 30 W, 30 Pa for 30 s was processed to improve the contact transparency.

**3. Low temperature transport measurement.** Quasi-four terminal measurements were performed by applying a small ac excitation current $I_{ac}$ and a dc bias current $I_{dc}$ and measuring the ac voltage $V_{ac}$. The frequency of the ac excitation is typically 30.9 Hz and a noise level of <10 nV is usually reached on the lock-in amplifier. Depending on the normal state resistance, $I_{ac}$ ranges between 0.2 nA and 1 nA to achieve high quality signals. The differential resistance can then be obtained $dV/dI=V_{ac}/I_{ac}$. All the measurements were carried out in a cryogen-free dilution refrigerator at a base temperature of ~10 mK with appropriate low-pass filters.

## 4. Characterization of Bi$_2$O$_2$Se

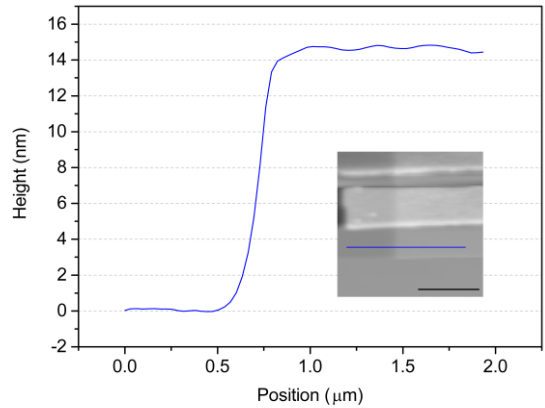

**Figure S2. Thickness of the nanoplate.** The thickness of the Bi$_2$O$_2$Se nanoplate used for the Josephson junction studied in this work was confirmed by atomic force microscopy (AFM). The image displays the extracted curve along the blue line on the AFM picture of the device (inset), showing the thickness of ~15 nm. Scale bar: 1 μm.

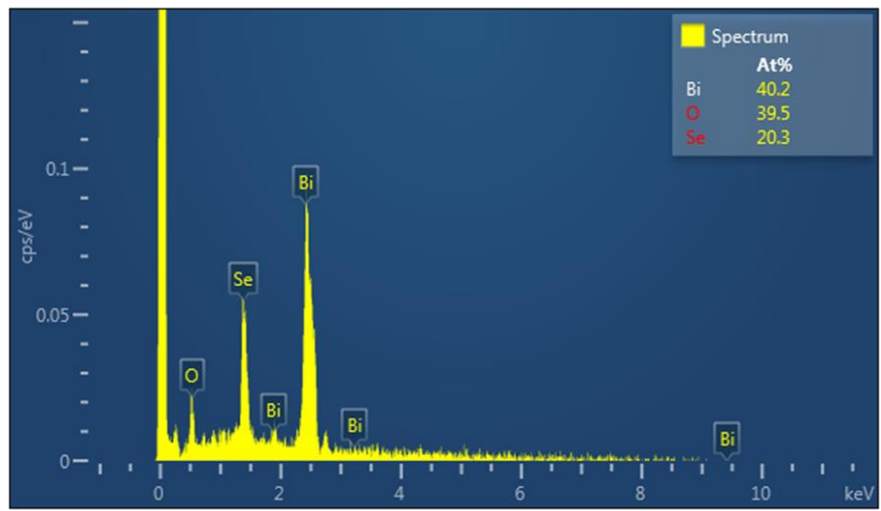

**Figure S3. EDS results.** The figure represents the energy dispersive spectroscopy (EDS) results of the as-grown nanoplates, confirming the atomic ratio of 2:2:1 for the Bi$_2$O$_2$Se phase.

## 5. Extraction of the supercurrent-density profile

In this subsection, we will describe the quantitative way we convert our measured interference patterns to the originating supercurrent-density profiles $J_s(x)$.

For a Josephson junction, the magnetic-field dependence of the maximum Josephson current (critical current) $I_c(B_z)$ is given by the Fourier transform of the supercurrent density $J_s(x)$ between the leads[3,4]. Here we assume that the current density varies only along the $x$ direction, the junction has a length $L$ in the $y$ direction, and the leads each have a length $L_{Al}$. Considering the flux focusing effect of the superconducting contacts, the effective length of the junction is $L_{eff}=L+L_{Al}$. Below we present the data measured at $V_g=0$ V for a larger magnetic field ($B_z$) range than Fig. 3a in the main text, and describe how $J_s(x)$ is retrieved. Note that the larger $B_z$ range allows a higher resolution of the edge supercurrent-density profile.

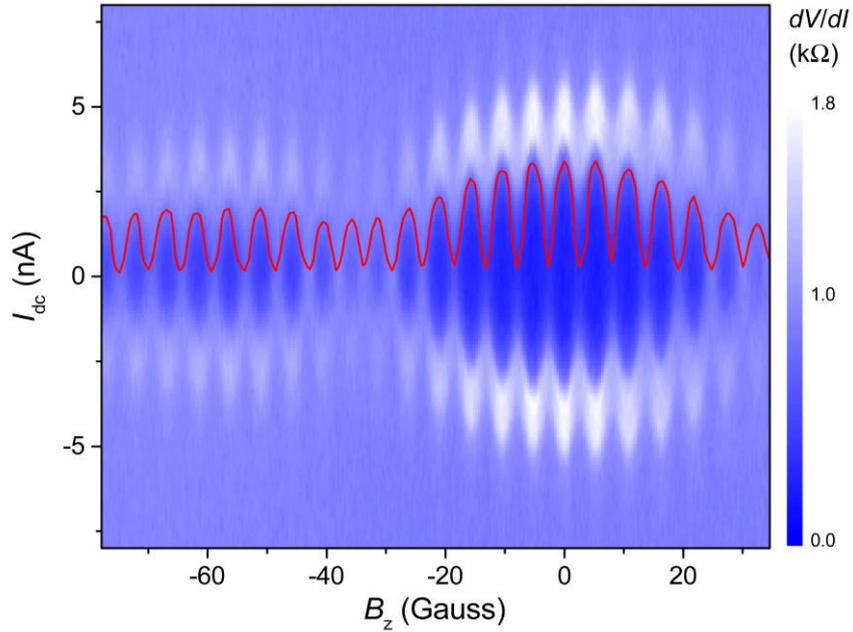

**Figure S4. Superconducting interference in a large magnetic field range.** Differential resistance d$V$/d$I$ as a function of perpendicular magnetic field $B_z$ and dc bias current $I_{dc}$ measured at $V_g=0$ V. The red curve illustrates $I_c(B_z)$ taken manually following the whitish envelope for an example.

Figure S4 shows the superconducting interference pattern measured at $V_g$=0 V. Clearly, the large scale envelope of $I_c(B_z)$ exhibits a node around $B_z$=-35 Gauss, resulting from the two-slit SQUID interference between the two edges imposed on the one-slit Fraunhofer interference of each edge, which indicates a finite width of the edge modes[5]. The red curve illustrates $I_c(B_z)$ taken manually following the whitish envelope, just for an example. But the experimentally observed $I_c(B_z)$ is the magnitude of the integration of $J_s(x)$. Thus, it is necessary to first recover the complex critical current $\Im_c(\kappa)$ in order to extract the supercurrent density from $I_c(B_z)$, where $\kappa=2\pi L_{eff} B_z/\Phi_0$ is the normalized magnetic field unit and $\Phi_0=h/2e$ the flux quantum ($h$ is the Planck constant and $e$ the elementary charge).

$$I_c(\kappa) = |\Im_c(\kappa)| = |\int_{-\infty}^{\infty} J_s(x) e^{i\kappa x} dx|$$

When we consider an even current density $J_{even}(x)$, representing a symmetric distribution, the problem will be reduced since the odd part of $e^{i\kappa x}$ vanishes from the integral, and we can obtain

$$\Im_c(\kappa) = I_{even}(\kappa) = \int_{-\infty}^{\infty} J_{even}(x) \cos(\kappa x) dx$$

According to the formula above, $\Im_c(\kappa)$ alternates between positive and negative values at each zero-crossing. Because the experimentally observed critical current $I_c(\kappa) = |\Im_c(\kappa)| = \sqrt{I_{enen}^2(\kappa) + I_{odd}^2(\kappa)}$ is dominated by $I_{even}(\kappa)$ except at its minima, approximately, $I_{even}(\kappa)$ can be obtained by multiplying $I_c(\kappa)$ with a flipping function that switches sign between adjacent lobes of the envelope function, as shown in Fig. S5.

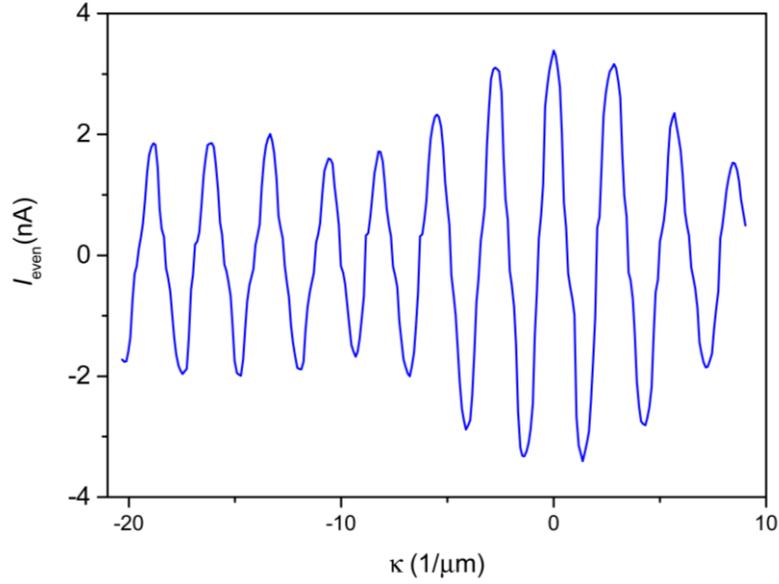

**Figure S5. The even current.** $I_{even}(\kappa)$ obtained by flipping sign for the even number of lobes of the red curve in Fig. S4.

Now suppose that there also exists a small but non-vanishing odd component $J_{odd}(x)$ besides the even current distribution. The odd part $I_{odd}(\kappa) = \int_{-\infty}^{\infty} J_{odd}(x)\sin(\kappa x)dx$ can then be approximated by interpolating between the minima of $I_c(\kappa)$, and flipping sign between lobes. The result is shown below in Fig. S6.

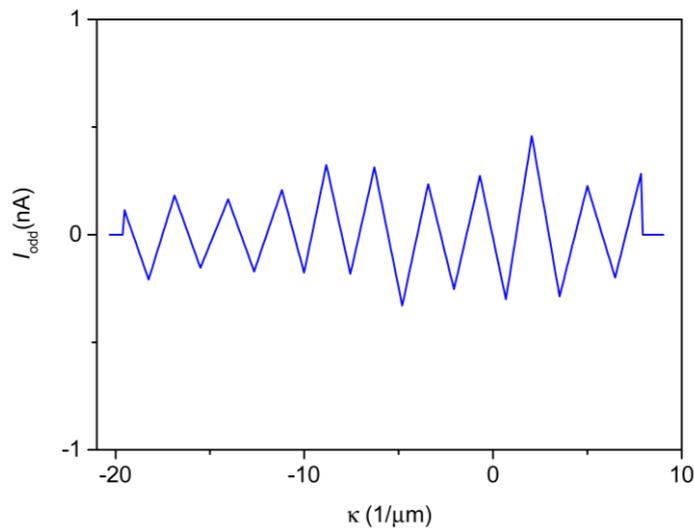

**Figure S6. The odd current.** The odd component $I_{odd}(\kappa)$ of the red curve in Fig. S4.

A Fourier transform of the resulting complex $\Im_c(\kappa)=I_{even}(\kappa)+iI_{odd}(\kappa)$, yields the supercurrent-density profile as shown in Fig. S7.

$$J_S(x) = |\frac{1}{2\pi}\int_{-W/2}^{W/2}\Im_C(\kappa)e^{-i\kappa x}d\kappa|$$

where $W$ is the width of the junction.

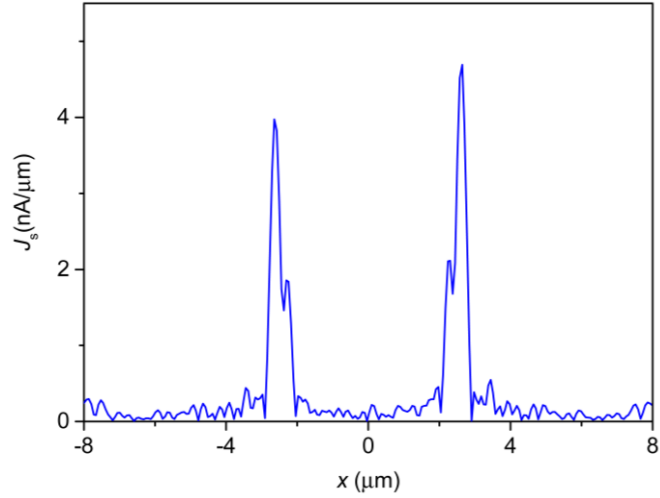

**Figure S7. Supercurrent density.** Supercurrent-density profile $J_s(x)$ obtained from the Fourier transform of the combination of Figs. S5 and S6.

The critical current $I_c(B_z)$ can, of course, be extracted from the superconducting interference pattern numerically using a program by setting a $dV/dI$ threshold counting from $I_{dc}=0$ towards high $I_{dc}$. Figures S8 and S9 display the results with a threshold of 650 Ω and 750 Ω, respectively. By a comparation between Figs. S7, S8d and S9d, we can see that how exactly the $I_c(B_z)$ curve is retrieved affects the details of the extracted $J_s(x)$. However, the overall features are reliable.

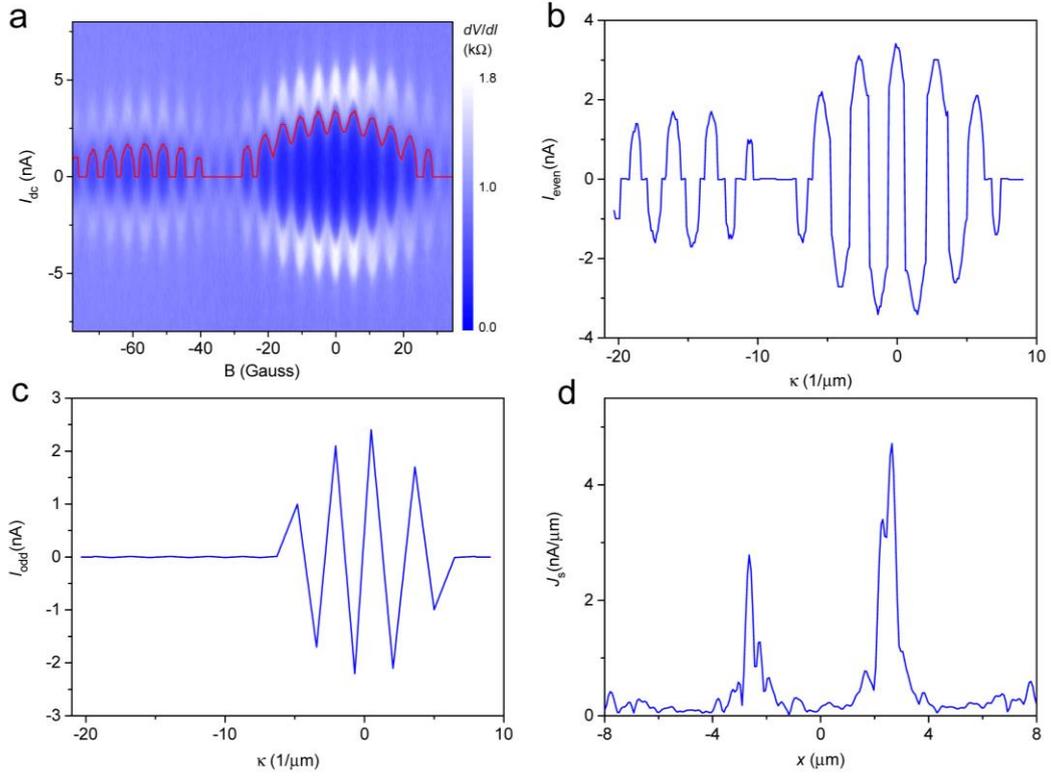

**Figure S8. Extraction of supercurrent density with a threshold of 650 Ω.** $I_c(B_z)$ was retrieved using a program by setting a *dV/dI* threshold of 650 Ω (red curve in **a**). **b-d**, $I_{even}(\kappa)$, $I_{odd}(\kappa)$ and $J_S(x)$, respectively.

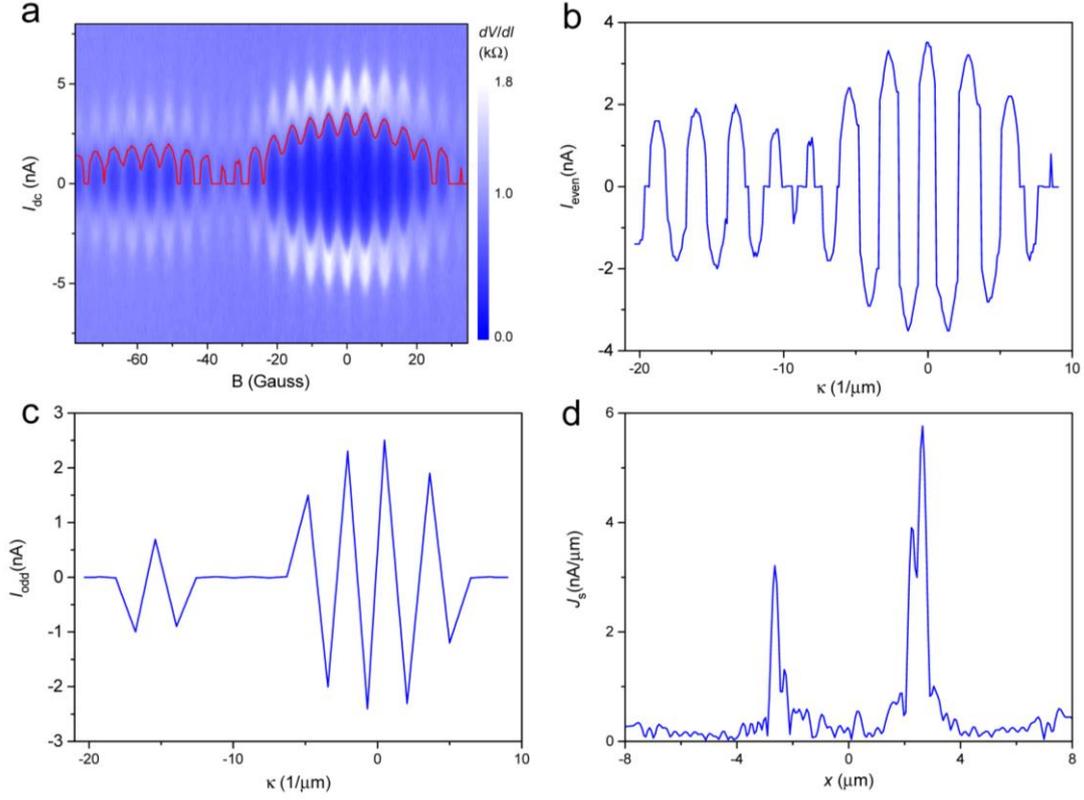

**Figure S9. Extraction of supercurrent density with a threshold of 750 Ω.** $I_c(B_z)$ was retrieved using a program by setting a *dV/dI* threshold of 750 Ω (red curve in **a**). **b-d**, $I_{even}(\kappa)$, $I_{odd}(\kappa)$ and $J_S(x)$, respectively.

Another noticeable feature of the supercurrent-density profiles in Figs. S7, S8d and S9d obtained from the large-range ($B_z$) interference pattern is that the supercurrent density presents a sharp peak at the very edge, and decays towards the center of the junction. This characteristic corroborates the interpretation of the edge channel width of ~570 nm in the main text which results from the extension of the electron waves from the edges into the bulk for Se-terminated top/bottom surface (from the side surfaces onto the top and bottom surfaces for Bi-terminated case), showing a decaying behavior.

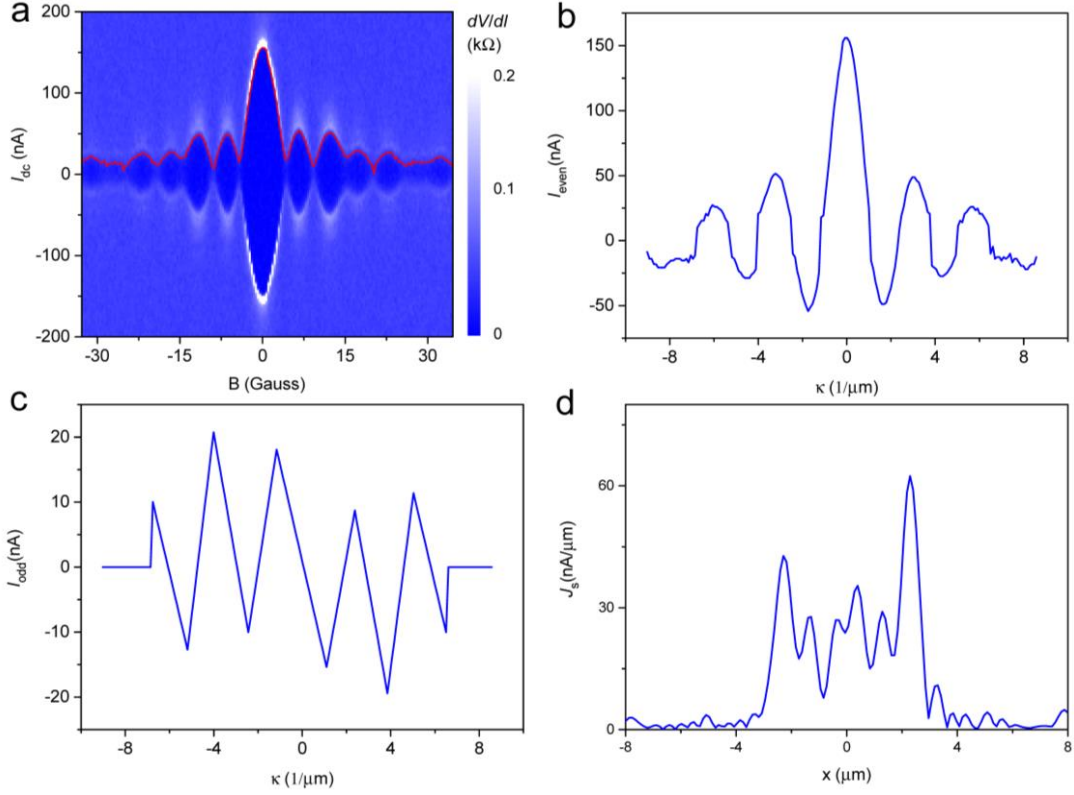

**Figure S10. Extraction of supercurrent density at $V_g$=40 V.** A $dV/dI$ threshold of 35 Ω was set in the program to retrieve $I_c(B_z)$, as shown by the red curve in **a** (the same data as in Fig. 3a in the main text for $V_g$=40 V). **b-d**, $I_{even}(\kappa)$, $I_{odd}(\kappa)$ and $J_S(x)$, respectively. **d** is also the same as in Fig. 3b in the main text for $V_g$=40 V.

## 6. Comparison between different devices

We measured several other devices fabricated on different $Bi_2O_2Se$ nanoplates. The most striking difference of these devices is the thickness of the nanoplate. We found that for thin devices, it is easy to realize the bulk to edge supercurrent transition. But the thicker the nanoplate, the harder to tune the critical supercurrent and thus the supercurrent transition. Below we show three typical junctions fabricated on nanoplates with a thickness of 22 nm, 42 nm and 64 nm, respectively.

To have a direct comparison between these devices, we normalized the change of the normal state resistance to the value at $V_g$=0 V, $[R-R(V_g=0\ V)]/R(V_g=0\ V)$, as shown in Fig. S11a. The inset displays an enlarged view of the red and blue curves. We can see that for the 22 nm-thick device, the resistance changes about 20 times from $V_g=0$ V to -20 V. But for the 42 nm- and 64 nm-thick devices, the resistance changes ~10% and ~2%, respectively. So, the thicker the nanoplate, the harder to tune the resistance, presumably due to the bulk carriers and/or screening of the top-layer carriers by the bottom-layer carriers since we applied a back-gate.

Figure S11b shows the 2D color map of the $dV/dI$ as a function of $V_g$ and $I_{dc}$ for the 22 nm-thick junction. Similar as the 15 nm-thick junction shown in the manuscript, the supercurrent can be tuned ON and OFF, at a more negative gate voltage. Similar bulk to edge supercurrent transition was also observed. In general, such transition can be achieved for devices fabricated on thin $Bi_2O_2Se$ nanoplates.

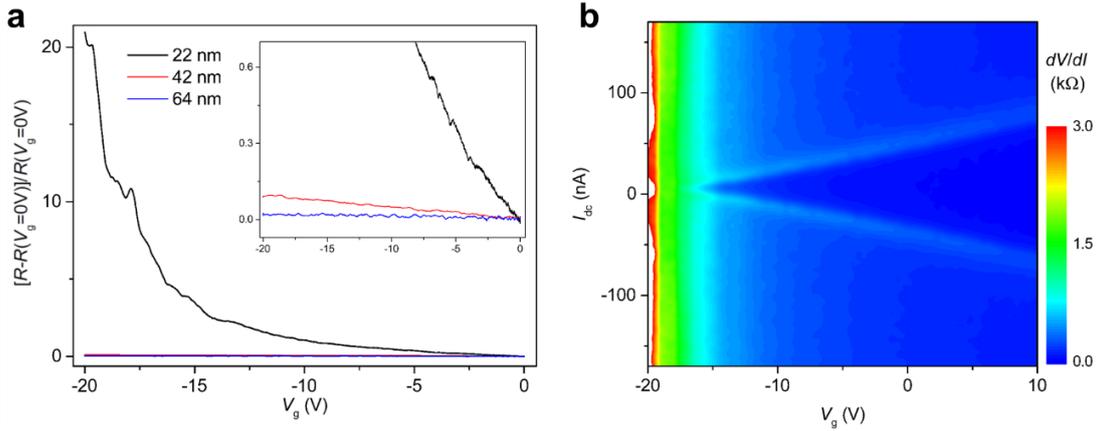

**Figure S11. Comparison between devices fabricated on nanoplates with different thickness. a,** $[R-R(V_g=0\ V)]/R(V_g=0\ V)$ as a function of gate voltage for Josephson devices fabricated on $Bi_2O_2Se$ nanoplates with a thickness of 22 nm, 42 nm, and 64 nm, respectively. **b,** $dV/dI$ as a function of $V_g$ and $I_{dc}$ for the 22 nm-thick junction.

At zero gate voltage, the normal state resistance is 152 Ω, 7.9 Ω and 8.3 Ω for the 22 nm, 42 nm and 64 nm thick devices, respectively. Accordingly, the width and length for these junctions are (2.9 μm, 300 nm), (4 μm, 160 nm) and (4.2 μm, 210 nm). The measurements were performed at ~10 mK. For the 15 nm-thick junction shown in the main text, the resistance at $V_g=0$ V is 812 Ω, and the dimensions are (4.6 μm, 300 nm).

Due to the quasi-four terminal configuration (Fig. 1b in the main text), the contact resistance is included in the normal state measurement. A direct comparation between these resistance values cannot reflect the resistance of the pure nanoplates, though, the trend shows that the bulk contribution is significant in the thick junctions.

Regarding the carrier density and mobility, we can apply the capacitor model to calculate the tunability of the density by a gate voltage. Using the thickness of $SiO_2$ of 300 nm, we get the density change of 1.4E12 cm$^{-2}$ from $V_g$=0 V to -20 V. To estimate the carrier density from the measured resistance, we need to assume that the mobility does not change. For Hall bar devices, we did find the drop of mobility following the decrease of density, from ~1.2E4 cm$^2$/Vs at 2E13 cm$^{-2}$ to ~5E3 cm$^2$/Vs at 1.2E13 cm$^{-2}$. However, the drop of mobility slows down at low density. Therefore, for the 42 nm-thick junction, which shows a resistance change of ~10%, we attribute to a change of density of ~10%. Consequently, the density at zero gate voltage is ~1.4E13 cm$^{-2}$. From the Hall bar measurement, we estimate a mobility of ~5E3 cm$^2$/Vs.

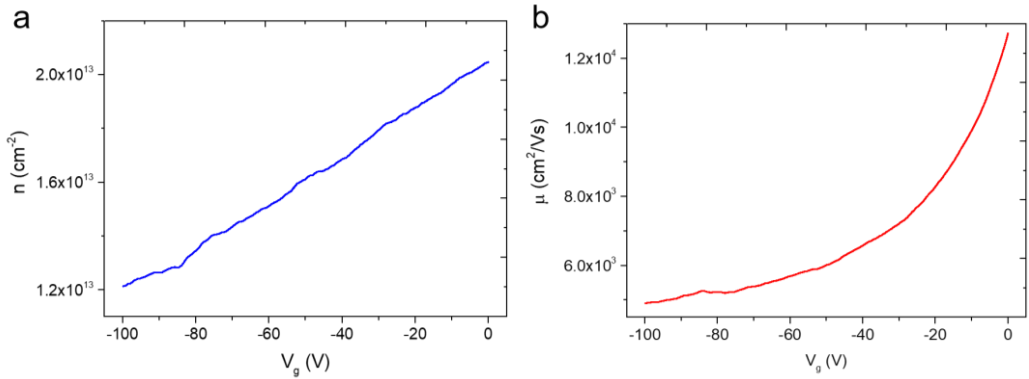

**Figure S12. Density and mobility measurement of a Hall bar device. a,** Electron density as a function of $V_g$. **b,** Mobility as a function of $V_g$.

However, this method does not apply to the 22 nm-thick junction, since the resistance changes ~20 times. Instead, for the 22 nm-thick junction, we assume that the Fermi level reaches the gap edge at $V_g$=-20 V since the resistance increases towards infinite, similar as the 15 nm-thick junction in the main text. Thus, the density is ~1.4E12 cm$^{-2}$ at $V_g$=0 V. For the 64 nm-thick junction, if the same treatment is applied as for the 42 nm-thick junction, a very high density of ~7E13 cm$^{-2}$ is expected, which we do not think is reliable. So, we only estimate the carrier density for the 22 nm and 42 nm thick junctions.